\documentclass[a4paper]{jpconf}
\usepackage{graphicx}

\newcommand {\nc} {\newcommand}
\nc {\bce}{\begin{center}}
\nc {\ece} {\end{center}}
\nc {\bit} {\begin{itemize}}
\nc {\eit} {\end{itemize}}
\nc {\beq} {\begin{eqnarray}}
\nc {\eeq} {\end{eqnarray}}
\nc {\ex} [1] {\ensuremath{^{#1}}}
\nc {\ve} [1] {\mbox{\boldmath $#1$}}
\nc {\dou} {$\Rightarrow$\,}
\nc {\flim} [2] {\mathop{\longrightarrow}\limits_{{#1}\rightarrow{#2}}}
  \nc {\eq} [1] {(\ref{#1})}
  \nc {\Eq} [1] {Eq.~(\ref{#1})}
  \nc {\Sec} [1] {Sec.~\ref{#1}}
  \nc {\fig} [1] {Fig.~\ref{#1}}
    \nc {\Ref} [1] {Ref.~\cite{#1}}
  \nc {\Sch} {Schr\"odinger }
  
\begin{document}
\title{Recent developments in the eikonal description of the breakup of exotic nuclei}

\author{\underline{P.~Capel}\ex{1}, F.~Colomer\ex{1}, H.~ Esbensen\ex{2}, T.~Fukui\ex{3}, R.~C.~Johnson\ex{4}, F.~M.~Nunes\ex{5}, K.~Ogata\ex{3}}

\address{\ex{1}Physique Nucl\'eaire et Physique Quantique C.P. 229, Universit\'e Libre de Bruxelles (ULB), B-1050, Brussels, Belgium\\
\ex{2}Physics Division, Argonne National Laboratory, Argonne, Illinois 60439, USA\\
\ex{3}Research Center for Nuclear Physics, Osaka University, Ibaraki, Osaka 567-0047, Japan\\
\ex{4}Department of Physics, University of Surrey, Guildford, England GU2 7XH, United Kingdom\\
\ex{5}National Superconducting Cyclotron Laboratory and Department of Physics and Astronomy, Michigan State University, East Lansing, Michigan 48824, USA}

\ead{pierre.capel@ulb.ac.be}

\begin{abstract}
The  study  of  exotic  nuclear  structures,  such  as  halo  nuclei,  is  usually  performed  through
nuclear reactions.
An accurate reaction model coupled to a realistic description of the projectile is needed to correctly interpret experimental data.
In this contribution, we briefly summarise the assumptions made within the modelling of reactions involving halo nuclei.
We describe briefly the Continuum-Discretised Coupled Channel method (CDCC) and the Dynamical Eikonal Approximation (DEA) in particular and present a comparison between them for the breakup of $^{15}$C on Pb at $68A$MeV.
We show the problem faced by the models based on the eikonal approximation at low energy and detail a correction that enables their extension down to lower beam energies.
A new reaction observable is also presented.
It consists of the ratio between angular distributions for two different processes, such as elastic scattering and breakup.
This ratio is completely independent of the reaction mechanism and hence is more sensitive to the projectile structure than usual reaction observables, which makes it a very powerful tool to study exotic structures far from stability.
\end{abstract}

\section{Introduction}
The development of radioactive-ion beams in the mid-80s has enabled us to study the structure of nuclei away from stability.
This has led to the discovery of exotic structures such as halo nuclei \cite{Tan96}.
Halo nuclei exhibit an extended matter radius compared to stable nuclei.
The unusual size of these neutron-rich nuclei is due to their small binding energy for one or two neutrons.
Thanks to this loose binding, the valence neutron(s) tunnel far into the classically forbidden region and form a diffuse halo around a tight and well-bound core.
The best known one-neutron halo nuclei are $^{11}$Be and $^{15}$C, which can be seen as a $^{10}$Be or $^{14}$C core, respectively, to which one neutron is loosely bound.
Archetypical two-neutron halo nuclei are $^6$He and $^{11}$Li, with an $\alpha$ and $^9$Li core, respectively.
%In addition to their halo, these nuclei also exhibit a Borromean structure, which means that although the three-body system core-n-n is bound, none of the two-body subsystems is (neither $^5$He nor $^{10}$Li exists, and the di-neutron is not stable in particle either).
%This is reminiscent of the Borromean rings, which consist of three rings entangled in such a way that when one of them is broken, the other two get loose, hence the name for these very exotic structures coined by Zhukov \etal.
Albeit much less probable, proton haloes are also possible; \ex{8}B is one of the nuclei that probably exhibit a one-proton halo.

Being located close to the drip lines, halo nuclei exhibit very short lifetimes, which impedes the use of regular spectroscopic techniques in their study.
To study these fascinating structures, we must then rely on indirect techniques such as elastic scattering or breakup.
In the latter, the halo dissociates from the core during the collision with a target, hence revealing the internal cluster structure of the projectile.
In order to infer correct nuclear-structure information from experimental data, an accurate theoretical description of the reaction, coupled to a realistic model of the nucleus under investigation is needed \cite{BC12}.
In this contribution, we review the theoretical framework of the usual reaction models and describe in particular the Continuum Discretise Coupled Channel model (CDCC) \cite{Kam86,YOM12} and the Dynamical Eikonal Approximation (DEA) (\Sec{model}) \cite{BCG05,GBC06}.
In \Sec{comparison}, we compare these two models on the case of the breakup of \ex{15}C on Pb at intermediate ($68A$MeV) and low ($20A$MeV) energies \cite{CEN12}.

Albeit precise, these models rely on inputs, such as optical potentials, which can significantly affect the theoretical predictions.
To avoid this dependence a new reaction observable has been recently proposed.
It consists of the ratio of angular distributions for two different processes, such as elastic scattering and breakup.
Taking this ratio significantly reduces the sensitivity to the reaction process and hence to the model inputs.
Accordingly this ratio is more sensitive to the projectile structure than usual reaction observables and can provide very accurate information about exotic nuclear structures.
The ratio method is presented in \Sec{ratio} before a brief summary in \Sec{summary}.

\section{Reaction modelling}\label{model}
\subsection{Theoretical framework}
The models of reactions involving one-nucleon halo nuclei usually rely on a three-body framework: a two-body projectile and a target.
The internal structure of the projectile ($P$) is described by the core ($c$)-fragment ($f$) Hamiltonian
\beq
H_0=T_r+V_{cf}(\ve{r}),
\label{e1}
\eeq
where $\ve{r}$ is the $c$-$f$ relative coordinate, $T_r$ is the kinetic energy term and $V_{cf}$ is a phenomenological potential that simulates the $c$-$f$ interaction, whose parameters are adjusted to reproduce the energy and quantum numbers of the low-energy levels of the projectile.
%\begin{figure}[h]
%\center
%\includegraphics[trim=4.5cm 17cm 6.5cm 0cm, clip=true, width=5.cm]{Tcf.pdf}
%\includegraphics[width=14pc]{name.eps}\hspace{2pc}%
%\caption{\label{f1}Jacobi set of coordinates which is used to describe the collision.}
%\end{figure}
The eigenstates of $H_0$ describe the $c$-$f$ relative motion
\beq
H_0\ \phi_{lm}(E,\ve{r})=E\ \phi_{lm}(E,\ve{r}),
\label{e2}
\eeq
where $l$ and $m$ are the $c$-$f$ relative orbital angular momentum and its projection, respectively.
$E$ is the $c$-$f$ relative energy.
The negative-energy states ($E<0$) are discrete; they correspond to the bound states of the projectile.
The positive energy states correspond to the continuum of the $c$-$f$ system, i.e. the states in which the halo has broken up from the core.

The target ($T$) is usually described as a structureless particle which interacts with the projectile constituents though optical potentials $V_{cT}$ and $V_{fT}$.
Within that framework, studying the reaction reduces to solving the following three-body \Sch equation
\beq
\left[T_R+H_0+V_{cT}+V_{fT}\right]\Psi(\ve{r},\ve{R})=E_T\ \Psi(\ve{r},\ve{R})
\label{e3}
\eeq
under the condition that the projectile is initially in its ground state $\phi_{l_0 m_0}(E_0)$
\beq
\Psi(\ve{r},\ve{R})\flim{Z}{-\infty}e^{iKZ+\cdots}\phi_{l_0,m_0}(E_0,\ve{r}),\label{e4}
\eeq
where the $Z$ axis has been chosen along the incoming beam and the $P$-$T$ momentum is related to the total energy $E_T$ of the system: $\hbar^2K^2/2\mu_{PT}+E_0=E_T$, with $\mu_{PT}$ the $P$-$T$ reduced mass.

Various models have been developed to solve this equation numerically.
In the following sections, we will describe two of them: the Continuum-Discretised Coupled Channel method (CDCC) and the Dynamical Eikonal Approximation (DEA).

\subsection{CDCC}
In CDCC, the three-body wave function $\Psi$ is expanded upon the complete set of the eigenstates of $H_0$:
\beq
\Psi(\ve r, \ve R)=\sum_{lm}\sum_{E_i<0}\chi_{lm}(E_i,\ve{R})\phi_{lm}(E_i,\ve{r})+\int_0^\infty \chi_{lm}(E,\ve{R})\phi_{lm}(E,\ve{r})\ dE.
\label{e5}
\eeq
To be tractable numerically, the integral over the positive $c$-$f$ energy must be discretised.
This is usually performed dividing the continuum into small bins and defining for each bin $i$ a normalised wave function
\beq
\tilde\phi_{i}(\ve r)=\frac{1}{N}\int_{E_i-\Delta E_i/2}^{E_i+\Delta E_i/2}w(E)\,\phi_{lm}(E,\ve{r})\ dE,
\label{e6}
\eeq
where the index $i$ defines not only the energy, but also the partial wave $lm$ of the bin state, $w$ is a weight function and $N$ is a normalisation coefficient.

With this discretisation, introducing the expansion \eq{e5} into \Eq{e3}, leads to a tractable set of coupled equations in the unknown coefficients $\chi_i$ of the expansion \eq{e5}
\beq
\left[T_R+E_i+V_{ii}\right]\chi_i+\sum_{j\ne i}V_{ij}\,\chi_j=E_T\ \chi_i,
\label{e7}
\eeq
where the couplings are due to the tidal force resulting from the difference of the action of the target $T$ upon the core $c$ and the halo $f$
\beq
V_{ij}=\langle\tilde\phi_i|V_{cT}+V_{fT}|\tilde\phi_j\rangle.
\label{e8}
\eeq

\subsection{Dynamical eikonal approximation}
The DEA is based on the eikonal approximation \cite{Glauber}, which states that at sufficiently high energy, the three-body wave function $\Psi$ does not deviate much from the incoming plane wave of \Eq{e4}.
Accordingly, this suggests to perform the following factorisation
\beq
\Psi(\ve r,\ve R)=e^{iKZ}\,\widehat\Psi(\ve r, \ve R),
\label{e9}
\eeq
where $\widehat\Psi$ varies smoothly with $\ve{R}$.
Introducing  \Eq{e9} into \Eq{e3}, and neglecting the second-order derivative of $\widehat\Psi$ in front of its first-order derivative, leads to the DEA equation \cite{BCG05,GBC06}
\beq
i\hbar v \frac{\partial}{\partial Z}\widehat{\Psi}(\ve{r},\ve{b},Z)
=[H_0-E_0+V_{cT}+V_{fT}]
\widehat{\Psi}(\ve{r},\ve{b},Z),\label{e10}
\eeq
where $v=\hbar K/\mu_{PT}$ is the initial $P$-$T$ relative velocity.

\Eq{e10} must be solved for each value of $\ve b$, the transverse component of the $P$-$T$ relative coordinate $\ve R$, with the condition $\widehat\Psi\flim{Z}{-\infty}\phi_{l_0,m_0}(E_0)$.
Being simpler to solve than the set of coupled equations \eq{e7}, it leads to a much faster numerical scheme than CDCC \cite{CEN12}.
However, it is valid only at sufficiently high energy, whereas CDCC is not restricted in energy.

\section{Coulomb breakup of $^{15}$C}\label{comparison}
\subsection{$^{15}$C+Pb at $68A$MeV}
To compare the two aforementioned models, we have performed numerical calculations for the breakup of \ex{15}C on Pb at $68A$MeV \cite{CEN12}, which correspond to the experimental conditions of \Ref{Nak09}.
\begin{figure}[h]
\center
\includegraphics[trim=1.5cm 18cm 7cm 1.7cm, clip=true, width=7.6cm]{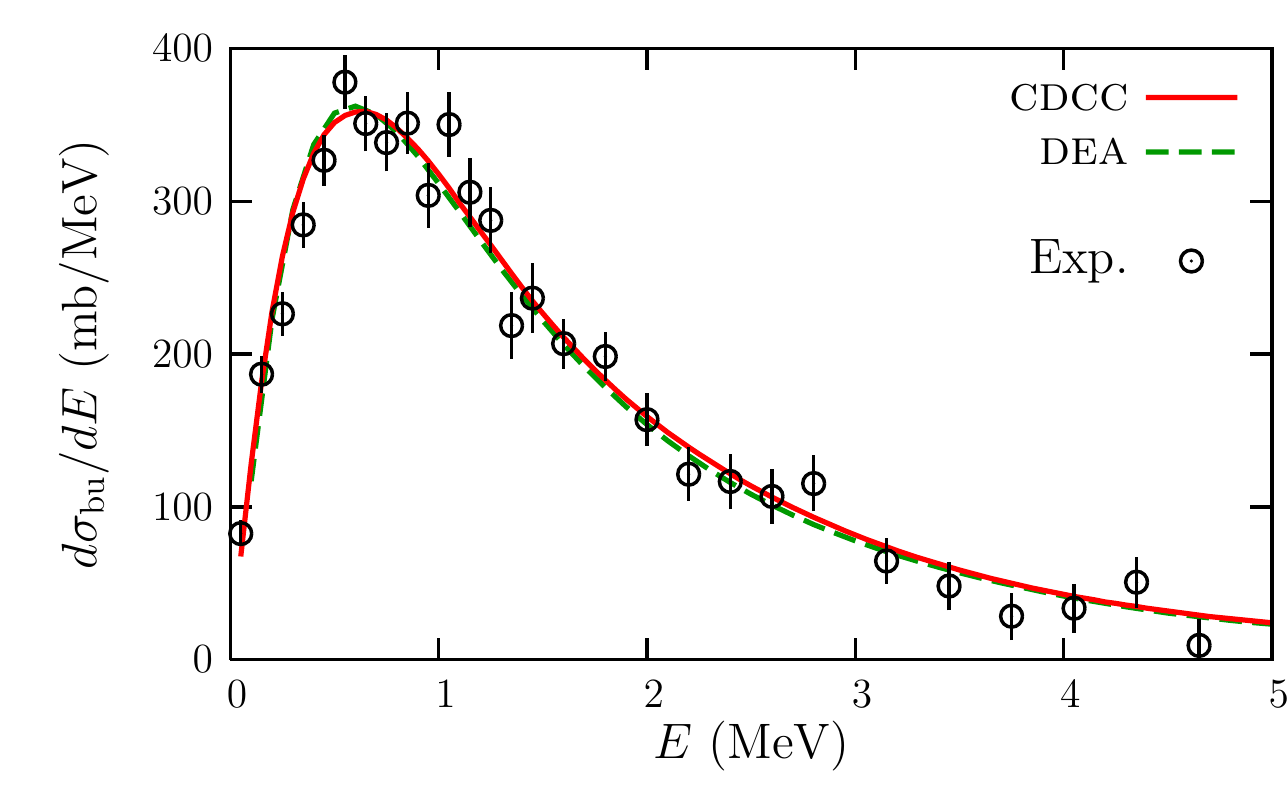}\hspace{5mm}
\includegraphics[trim=2.5cm 18cm 7cm 1.7cm, clip=true, width=7cm]{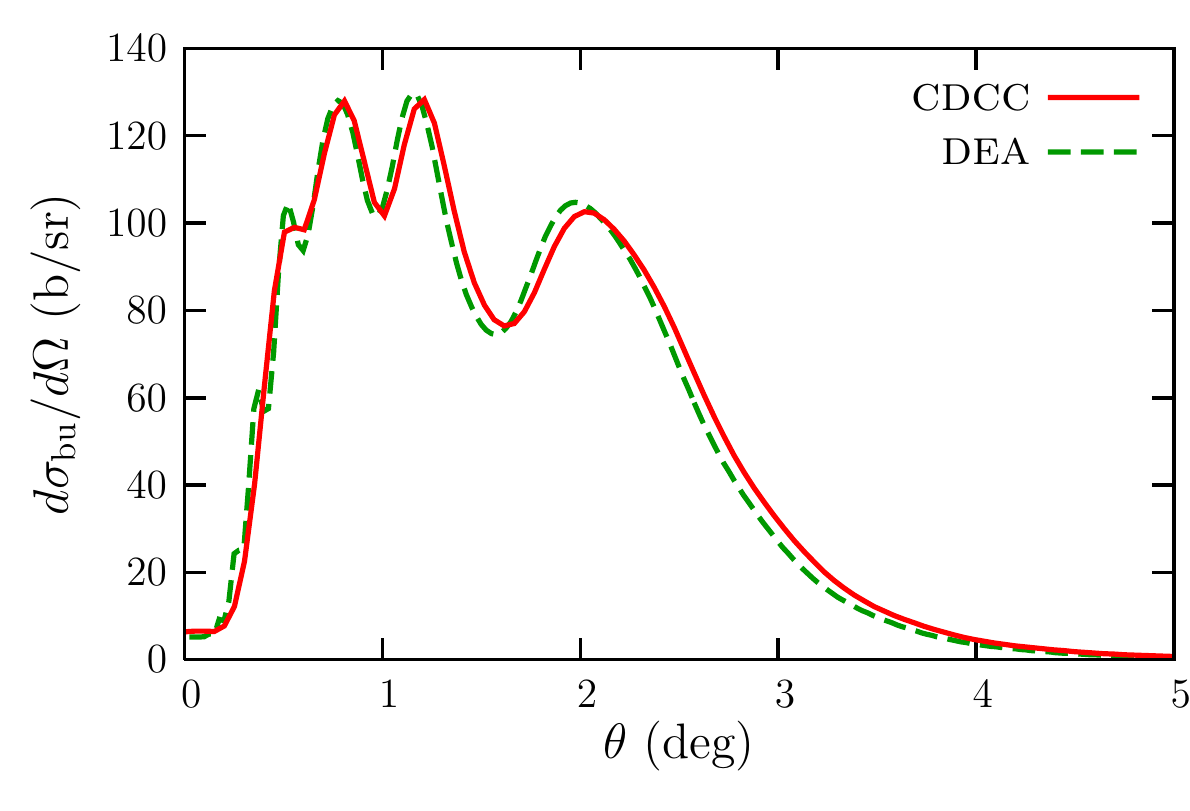}
\caption{\label{f2}Comparison of the CDCC (solid line) and DEA (dashed line) reaction models on the breakup of \ex{15}C on Pb at $68A$MeV: energy distribution (left) and angular distribution (right).
Calculations are from \Ref{CEN12}, data are from \Ref{Nak09}.}
\end{figure}
As can be seen in \fig{f2}, both models provide nearly identical breakup cross sections.
They lead to the same energy distribution (\fig{f2}, left) and a mere tiny shift is observed in the angular distribution (\fig{f2}, right).
This shows that although based on different assumptions, both models provide the same description of the reaction process, and hence lead to the same conclusions about the projectile structure.
In particular, the excellent agreement between the calculations and the experimental data of \Ref{Nak09} confirms the halo structure of \ex{15}C.

\subsection{$^{15}$C+Pb at $20A$MeV}
\begin{figure}[h]
\center
\includegraphics[width=7cm]{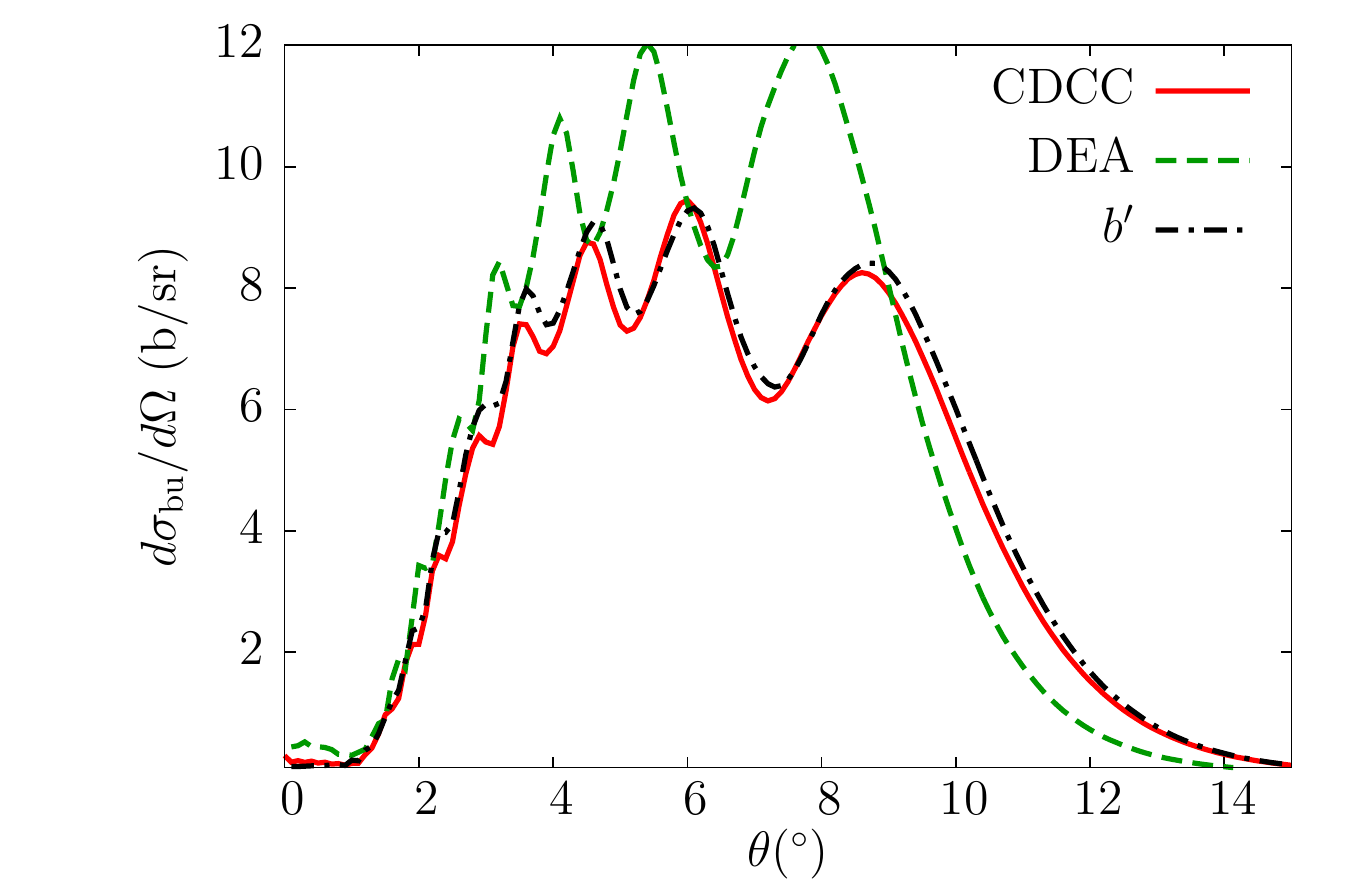}
\caption{\label{f3}Angular distribution computed for the breakup of \ex{15}C on Pb at $20A$MeV using CDCC (solid line) and DEA (dashed line).
A semiclassical correction (dash-dotted line) corrects perfectly the DEA \cite{FOC14}.}
\end{figure}

As mentioned above, the eikonal approximation is only valid at high energy \cite{Glauber}.
To see what happens when the beam energy is lowered, we have repeated the calculations at $20A$MeV \cite{CEN12}.
As illustrated in \fig{f3}, the DEA predicts a breakup cross section too large with an angular distribution which is too forward focused compared to CDCC, i.e.\ the slight shift observed in \fig{f2} (right) increases at low energy.
As analysed in \Ref{CEN12}, this is due to the lack of Coulomb deflection of the eikonal approximation.
Indeed, the factorisation \eq{e9} somehow forces the projectile to move forward along a straight line, leading it into the high-field zone of the target.
The predicted breakup cross section is hence too large and focused too forward in angle as compared to the case in which the Coulomb deflection off the target is properly taken into account.
This does not happen within CDCC, which naturally includes this deflection.

Analysing this effect in more detail in \Ref{FOC14}, we have observed that a simple semi-classical approximation \cite{BW81,BD04} corrects efficiently the eikonal framework for its lack of Coulomb deflection.
In that correction, the norm of the transverse component $\ve b$ of $\ve R$ is replaced by the distance of closest approached $b'$ between the projectile and the target on a classical Coulomb trajectory.
The resulting DEA calculation is now in perfect agreement with the CDCC calculation (see dash-dotted line in \fig{f3}). This shows that the range of validity of the DEA can be reliably extended down to low energy

\section{Ratio method}\label{ratio}
Albeit accurate, the aforementioned models rely on several inputs, including the optical potentials used to simulate the interaction between the projectile components and the target.
As shown in \Ref{CGB04}, the uncertainty related to this interaction can climb up to a factor 2 in the breakup cross section.
This is especially true for $V_{cT}$, which is usually poorly known since the core of the nucleus is often itself radioactive.

To circumvent this problem, a new reaction observable has been suggested \cite{CJN11}.
It consists of the ratio of angular distributions for different processes, e.g. elastic scattering and breakup.
Within the Recoil Excitation and Breakup model (REB), these distributions elegantly factorise into an elastic scattering cross section for a pointlike projectile and a form factor that accounts for the extension of the halo \cite{JAT97}:
\beq
\frac{d\sigma_{\rm el}}{d\Omega}&=&|F_{00}|^2\left(\frac{d\sigma}{d\Omega}\right)_{\rm pt}
\label{e11}
\eeq
for the elastic scattering cross section, with the form factor
\beq
|F_{00}|^2=\frac{1}{2l_0+1}\sum_{m_0} \left|\int |\phi_{l_0m_0}(E_0,\ve r)|^2e^{i\ve{Q\cdot r}}d\ve{r}\right|^2,
\label{e12}
\eeq
where $\ve{Q}\propto(\ve{K}-\ve{K'})$ is proportional to the exchanged momentum.
For breakup at energy $E$ in the $c$-$f$ continuum, one obtains \cite{CJN13}
\beq
\frac{d\sigma_{\rm bu}}{dE d\Omega}&=&
|F_{E0}|^2\left(\frac{d\sigma}{d\Omega}\right)_{\rm pt},
\label{e13}
\eeq
with the form factor
\beq
|F_{E0}|^2=\frac{1}{2l_0+1}\sum_{m_0}\sum_{lm} \left|\int \phi_{lm}(E,\ve r)\, \phi_{l_0m_0}(E_0,\ve r)\, e^{i\ve{Q\cdot r}}d\ve{r}\right|^2.
\label{e14}
\eeq
Since the pointlike cross section is the same in both factorisations \eq{e11} and \eq{e13}, taking the ratio of these angular distributions should remove most of the angular dependence and lead to an observable that depends only on the projectile wave functions.
However, the REB is based on two simplifying assumptions: it assumes the adiabatic approximation and it neglects $V_{fT}$.
To check these predictions we perform DEA calculations, which accounts for the dynamics of the projectile, includes the $f$-$T$ interaction, and is in excellent agreement with experiment \cite{GBC06}. 

Various tests have shown that it was best to consider the following ratio \cite{CJN13}
\beq
d\sigma_{\rm bu}/d\sigma_{\rm sum}&\stackrel{(REB)}{=}&|F_{E0}|^2
\label{e15}
\eeq
with the summed cross section corresponding to all elastic, inelastic and breakup processes
\beq
\frac{d\sigma_{\rm sum}}{d\Omega}&=&\frac{d\sigma_{\rm el}}{d\Omega}+\frac{d\sigma_{\rm inel}}{d\Omega}+\int \frac{d\sigma_{\rm bu}}{dEd\Omega} dE.
\label{e16}
\eeq

\begin{figure}[h]
\center
\includegraphics[trim=2.5cm 18.4cm 6.5cm 2cm, clip=true, width=8.9cm]{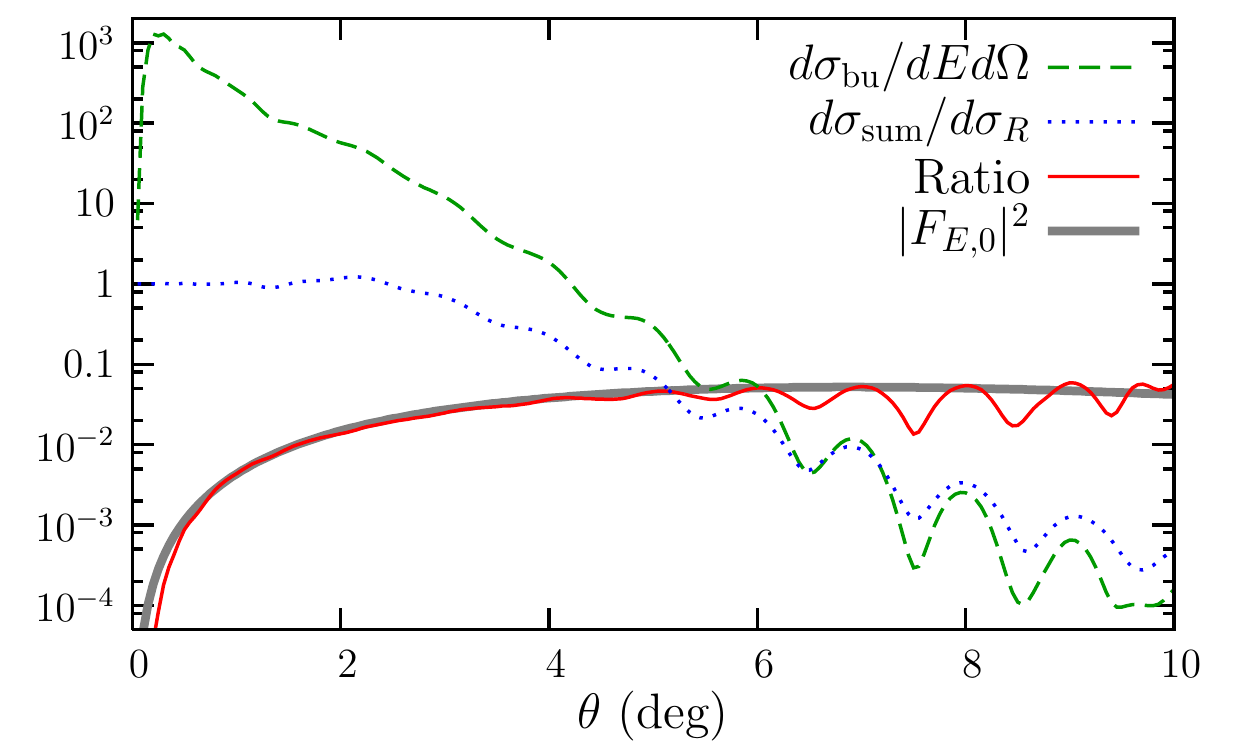}\hspace{2mm}
\includegraphics[width=6.7cm]{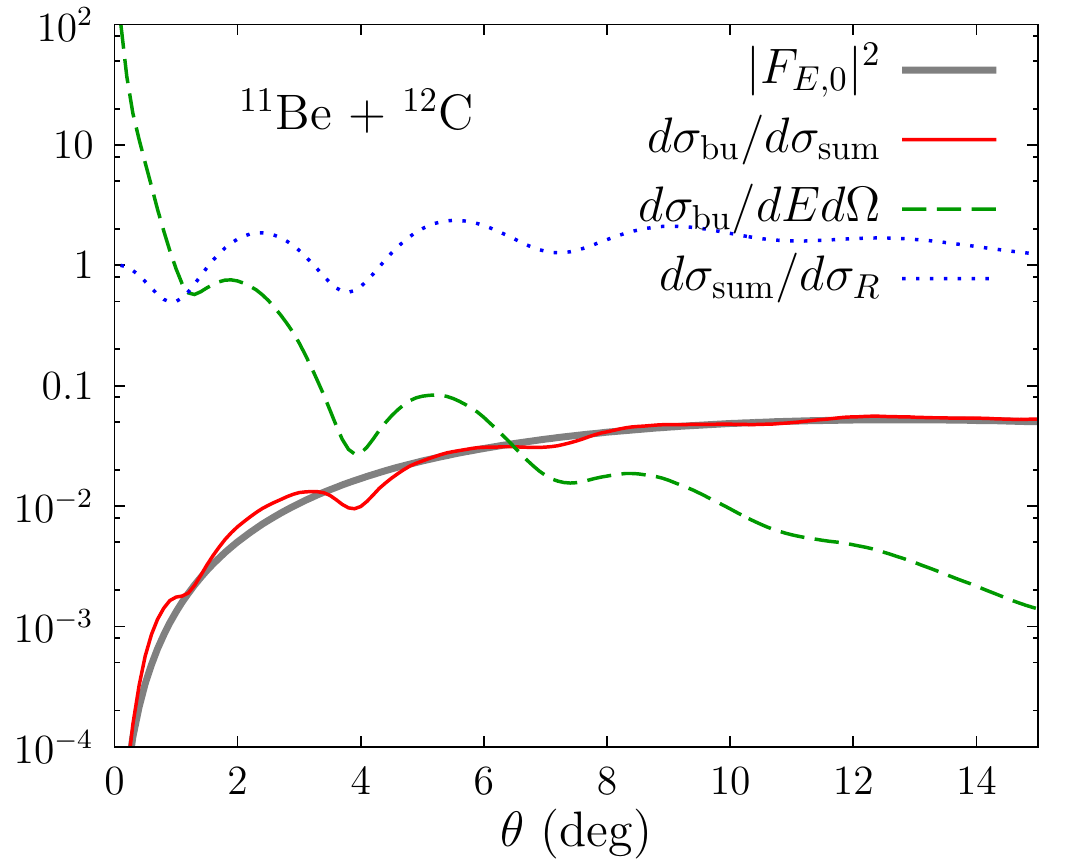}
\caption{\label{f4}Ratio method illustrated for the collision of \ex{11}Be on Pb at $69A$MeV (left) and on C at $67A$MeV (right).
Although the reaction process is very different in each reaction, the ratio~\eq{e15} are nearly identical and in excellent agreement with the REB prediction.}
\end{figure}

The results of these tests are displayed in \fig{f4} for \ex{11}Be impinging on Pb at $69A$MeV (left) and on C at $67A$MeV (right) \cite{CJN11,CJN13}. This figure shows the breakup cross section \eq{e13} (expressed in b/MeV sr, dashed lines), the summed cross section \eq{e16} (as a ratio to Rutherford, dotted lines), and their ratio \eq{e15} (solid lines).
We observe that although both collisions are dominated by very different processes and that, accordingly, their cross sections are very different, the ratio of the DEA cross sections is very smooth, nearly free from any oscillatory pattern, confirming that the ratio removes most of the dependence on the reaction process.
Moreover, in both cases it is in excellent agreement with the REB form factor \eq{e14} (thick grey line), showing that the ratio provides information about the projectile structure free from any reaction artefact, contrarily to usual reaction observables.
The ratio is especially sensitive to the $c$-$f$ binding energy and the partial wave of the projectile ground state \cite{CJN11,CJN13}.

\begin{figure}[h]
\center
\includegraphics[width=7.9cm]{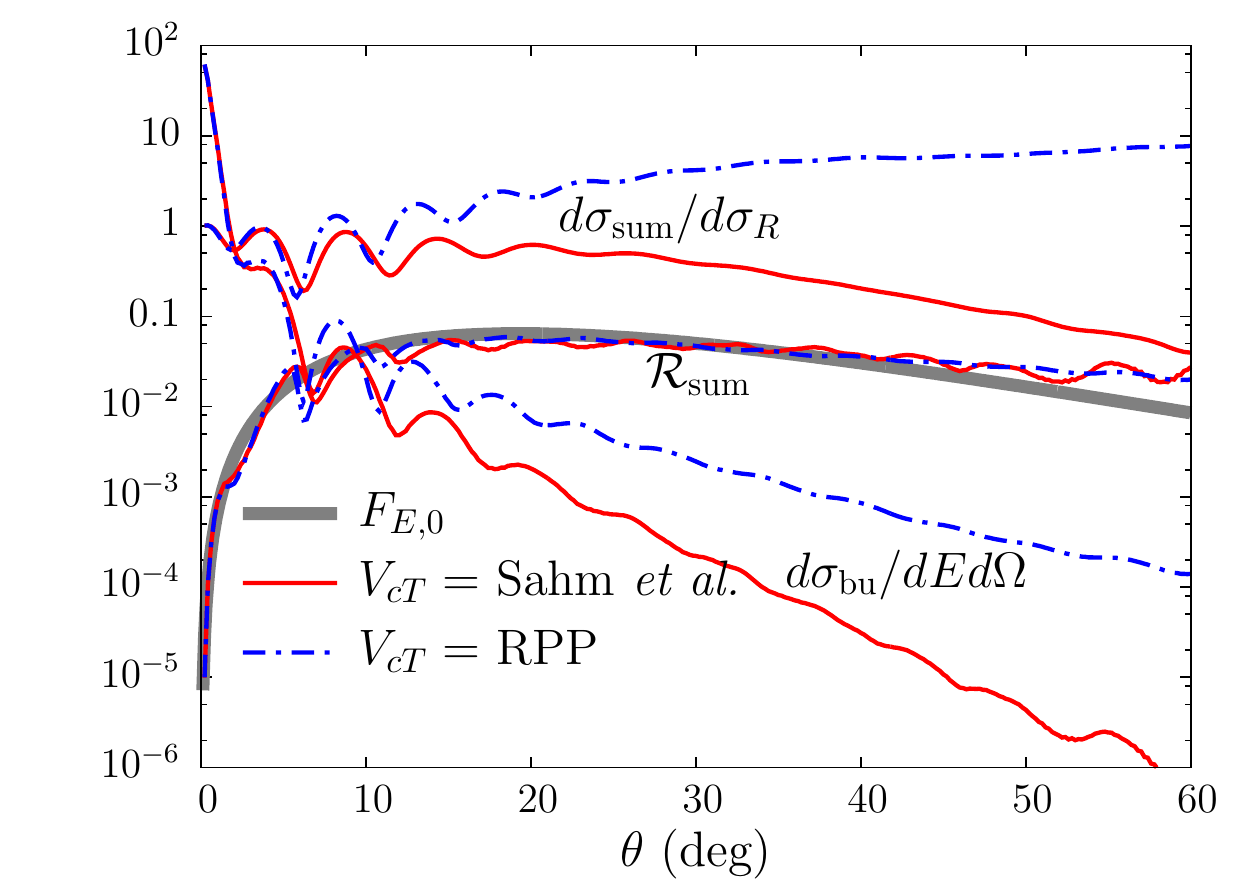}
\caption{\label{f5}Extension of the ratio method down to low energy ($20A$MeV) for \ex{11}Be on C.
Results obtained with different potentials (solid lines and dash-dotted lines) lead to identical ratios.}
\end{figure}

More recently, we have tested the validity of the ratio down to low energy.
Very surprisingly our calculations have shown that the method remains valid at low energy, especially on light targets \cite{CCN15}.
This analysis is illustrated in \fig{f5} for \ex{11}Be impinging on C at $20A$MeV.
Here again the ratio~\eq{e15} removes most of the angular dependence leading to an observable in excellent agreement with the REB prediction, even though this beam energy is outside the range of validity of that model.
The figure shows also results obtained with different $V_{cT}$.
As observed in \Ref{CGB04}, they produce significantly different cross sections.
However, the corresponding ratios are nearly identical.
These preliminary results confirm the power of the ratio method in the analysis of the nuclear structure away from stability.

\section{Summary}\label{summary}
Nuclear reactions like elastic scattering and breakup are useful tools to study exotic nuclear structure, such as halo nuclei \cite{Tan96}.
Most current models are based on a three-body framework: a two-body projectile impinging on a structureless target.
Various techniques have been developed to solve the corresponding three-body \Sch equation.
In this contribution, we have described two of them: CDCC \cite{Kam86} and DEA \cite{BCG05}.
Their comparison shows that at intermediate energy, they provide similar results \cite{CEN12}.
At lower energy, the eikonal approximation is no longer valid.
However a simple semiclassical correction enables us to extend its domain of validity down to low energies \cite{FOC14}.

To avoid the dependence of the reaction models to uncertain inputs such as optical potentials, a new reaction observable is suggested.
It consists of the ratio of  angular distributions for two different processes such as breakup and elastic scattering, which is predicted by the REB model \cite{JAT97} to depend only on the projectile wave functions.
Precise reaction calculations confirm this hypothesis and show the high sensitivity of the ratio to the projectile structure compared to regular reaction observables \cite{CJN13}.
This suggests the ratio method to be a very powerful tool to study the structure of nuclei far from stability.
Hopefully, this theoretical prediction will be confirmed by actual experimental data.

\section*{Acknowledgement}
This research was supported in part by the Research Credit No.~19526092 of the Belgian Funds for Scientific Research F.R.S.-FNRS, the National Science Foundation under Grant No. PHY-1403906, the Department of Energy, Office of Science, Office of Nuclear Physics under award No.\ DE-FG52-08NA28552.
This paper presents research results of the Belgian Research Initiative on eXotic nuclei
(BRIX), program P7/12, on interuniversity attraction poles of the Belgian Federal Science Policy Office.

\end{document}